\documentclass[twocolumn,showpacs,floatfix,superscriptaddress,amsmath,amssymb,prb]{revtex4}
\usepackage{amsmath}
\usepackage{mathrsfs}
\usepackage{txfonts}
\usepackage{amssymb}
\usepackage{graphicx}
\usepackage{hyperref}
\usepackage{ulem}
 \usepackage{overpic}
 \usepackage{psfrag}
 \newcommand{\be}{\begin{equation}}
\newcommand{\ee}{\end{equation}}

\begin{document}
\title{Ground-State Fidelity and Kosterlitz-Thouless Phase Transition for Spin $1/2$ Heisenberg Chain with Next-to-the-Nearest-Neighbor Interaction}
\author{Hong-Lei Wang, Ai-Min Chen, Bo Li and Huan-Qiang Zhou }

\affiliation{Centre for Modern Physics and Department of Physics,
Chongqing University, Chongqing 400044, The People's Republic of
China}

\begin{abstract}
The Kosterlitz-Thouless transition for the spin $1/2$ Heisenberg
chain with the next-to-the-nearest-neighbor interaction is
investigated in the context of an infinite matrix product state
algorithm, which is a generalization of the infinite time-evolving
block decimation algorithm [G. Vidal, Phys. Rev. Lett. \textbf{98},
070201 (2007)] to accommodate both the next-to-the-nearest-neighbor
interaction and spontaneous dimerization. It is found that, in the
critical regime, the algorithm automatically leads to infinite
degenerate ground-state wave functions, due to the finiteness of the
truncation dimension. This results in \textit{pseudo} symmetry
spontaneous breakdown, as reflected in a bifurcation in the
ground-state fidelity per lattice site. In addition, this allows to
introduce a pseudo-order parameter to characterize the
Kosterlitz-Thouless transition.

\end{abstract}


\maketitle

\section{introduction}

Quantum phase transitions (QPTs)~\cite{qpt} is one of the most
intriguing research subjects in condensed matter physics. A QPT
occurs at absolute zero due to quantum fluctuations in a variety of
quantum many-body systems. In the conventional
Landau-Ginzburg-Wilson paradigm, a phase transition accompanies
spontaneous symmetry breaking (SSB)~\cite{anderson,coleman} that is
characterized by a local order parameter. However, it now becomes
clear that not all QPTs fall into this category~\cite{wen}. In fact,
topological phase transitions do not involve any SSB. This even
dates back to the Kosterlitz-Thouless (KT) transition~\cite{kt},
first discovered for two-dimensional classical $XY$ model. Actually,
the KT transion is ubiquitous in one-dimensional quantum systems. It
describes the instability of the Luttinger liquid under a marginal
perturbation. Normally, it is not an easy task to determine whether
or not the KT transition occurs in a specific system, because there
are pathological problems to analyze the KT transition numerically.
One of these problems is that the finite size scaling
technique~\cite{fss}, which is successful for second order
QPTs~\cite{qpt,wen}, can not be applied to the KT
transition~\cite{np}, since there are logarithmic corrections from
the marginal perturbation.

Recently, a novel approach to QPTs in quantum many-body lattice
systems has been put
forward~\cite{zp,zhou,zov,fidelity1,fidelity2,fidelity3,rams}, which
is based on fidelity, a measure of quantum state distinguishability,
in quantum information science. As argued in Refs.~\cite{zhou,zov},
the ground-state fidelity per lattice site is able to capture
quantum criticality underlying many-body physics in condensed
matter. This fact, combining with a practical means to compute the
ground-state fidelity per lattice site for infinite-size quantum
lattice systems, make it practical to investigate critical phenomena
in quantum many-body systems. Coincidentally,  recent developments
in the context of the tensor network (TN) algorithms for
translation-invariant quantum lattice systems offer such a practical
means. Here, we mention the infinite matrix product state (iMPS)
algorithm~\cite{vidal1}  in one spatial dimension and the infinite
projected entangled-pair states (iPEPS)~\cite{PEPS} in two or higher
spatial dimensions.  These algorithms exploit the translation
invariance of the system and parallelizability of a TN
representation of quantum many-body wave functions, which provides
an efficient way to classically simulate quantum many-body lattice
systems.

In a previous work~\cite{zh}, we have succeeded in applying the
fidelity per site approach to study the KT transition in both the
one-dimensional spin $1/2$ XXZ model and the spin 1 XXZ model with
uniaxial single-ion anisotropy. This results in the introduction of
a novel concept-{\it pseudo} SSB, offering a novel perspective to
understand the KT transition, in the conventional
Landau-Ginzburg-Wilson paradigm, from the iMPS representation. As
such, it resolved the controversy regarding whether or not the
ground-state fidelity is able to detect the KT
QPTs~\cite{fidelity3}. However, more work is needed to clarify if
such an approach is applicable to the KT transition in other quantum
many-body lattice systems.


 \begin{figure}[!htp]
  \begin{center}
\includegraphics[angle=90,width=2.8in]{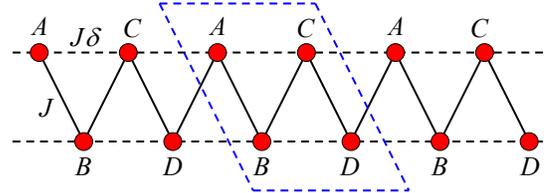}
\end{center}
\vspace{-0.3 cm}
 \caption{(Color online)
A sketch for the zigzag chain with the nearest-neighbor (NN) and
next-to-the nearest-neighbor (NNN) interactions J and J$\delta$. The
parameter $\delta$ represents the ratio between the NNN coupling and
the NN coupling. The dash-line box indicates the unit cell.}
  \label{Zigzag}
 \end{figure}

In this paper, we propose an iMPS algorithm, a generalization of the
infinite time-evolving block decimation algorithm~\cite{vidal1},
which allows us to accommodate both the next-to-the-nearest-neighbor
interaction and spontaneous dimerization. It is found that, in the
critical regime, the algorithm automatically leads to infinite
degenerate ground-state wave functions, due to the finiteness of the
truncation dimension. This results in \textit{pseudo} symmetry
spontaneous breakdown, as reflected in a bifurcation in the
ground-state fidelity per lattice site. In addition, this allows to
introduce a pseudo-order parameter to characterize the KT
transition, which  must be scaled down to zero in order to be
consistent with the Mermin-Wagner theorem~\cite{mw}.

\section{Matrix product state algorithm on an infinite-size one-dimensional lattice}
Suppose the model Hamiltonian takes the form: $H =\sum _i h^{[i,~
i+1,~i+2]}$, with $h^{[i,~i+1,~i+2]}$ being the sum of the
nearest-neighbor and the next-to-the-nearest-neighbor three-body
Hamiltonian density. A realization of such a model Hamiltonian in
the zigzag chain is shown in Fig.~\ref{Zigzag}. To take account of a
possible dimerization arising from the competition between quantum
fluctuations and geometric frustration, we choose four sequential
sites $(A,B,C,D)$ as a unit cell. Starting with a randomly chosen
initial state $|\Psi (0)\rangle $, which is not orthogonal to the
genuine ground state, a ground-state wave function can be computed
by the imaginary time evolution $ |\Psi_g\rangle = \exp(- H \tau)
|\Psi (0)\rangle / ||\exp(- H \tau) |\Psi (0)\rangle || $ with
$\tau\rightarrow\infty$. To realize the imaginary time evolution
operation, the imaginary time $\tau$ is divided into many small
slices $\delta \tau=\tau/N$ to approximate the continuous time
evolution by a sequence of small gates. Meanwhile, the time
evolution operator is expanded to a product of evolution operators
acting on three sites: $U_3 = \exp(-h^{[i,~i+1,~i+2]} \delta \tau )$
with $\delta \tau <<1$, as follows from the Suzuki-Trotter
decomposition~\cite{suzuki}.
Any wave function admits an iMPS representation in a canonical form:
attached to each site a four-index tensor $\Gamma^{slr}_{n}$ and
each bond a diagonal matrix $\lambda_n$, where $n=A,B,C \;{\rm and}
\;D$. Here, $s$ is a physical index, $s=1,\cdots,d$, with $d$ being
the dimension of the local Hilbert space. $l$ and $r$ denote the
bond indices, $l,r=1,\cdots, \chi$, with $\chi$ being the truncation
dimension. For simplicity, we use $\Gamma_{n}$  instead of
$\Gamma^{slr}_{n}$ in the text bellow.
%

 \begin{figure}[!htp]
 \begin{center}
\includegraphics[angle=90,width=3in]{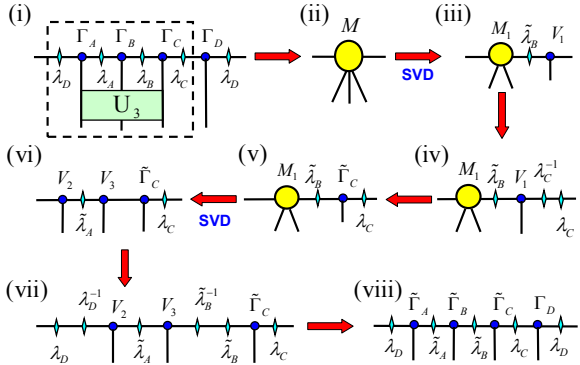}
\end{center}
 \caption{(Color online)
The  updating procedure for the tensors $\Gamma_n$ and the diagonal
matrices $\lambda_n$ ($n=A,B,C \;{\rm and}\; D$): (i) Apply the
three-site gate $U_3$. one needs to update the tensors $\Gamma_A$,
$\Gamma_B$, $\Gamma_C$, and the diagonal matrices $\lambda_A$,
$\lambda_B$ ; (ii) Contract the tensors inside the dash-line box in
(i) into a a single tensor $M$; (iii) Reshape $M$ into a matrix and
perform a singular value decomposition (SVD); (iv) Insert the
identity resolution on the right hand side; (v) Update the tensor
${\tilde\Gamma}_C$; (vi) Perform an SVD for the matrix contracted
from $M_1$ and ${\tilde \lambda}_B$; (vii) Insert the identity
resolution; (viii) All the tensors ${\tilde\Gamma}_A$,
${\tilde\Gamma}_B$, ${\tilde\Gamma}_C$, and the diagonal matrices
${\tilde\lambda}_A$ and ${\tilde\lambda}_B$ are updated.}
  \label{SVD}
 \end{figure}

The updating procedure for the tensors $\Gamma_{n}$ and the diagonal
matrices $\lambda_n$ ($n=A,B,C \;{\rm and}\; D$) in the iMPS
representation under the action of the three-site gate $U_3$  is
visualized in Fig.~\ref{SVD}: (i) Apply the three-site gate $U_3$
onto the iMPS tensors $\Gamma_A$, $\Gamma_B$ and $\Gamma_C$. The
tensors involved in this action is shown in the dash-line box. (ii)
Contract the tensors $\lambda_D$, $\Gamma_A$, $\lambda_A$,
$\Gamma_B$, $\lambda_B$, $\Gamma_C$ and $\lambda_C$ into a single
tensor and reshape the tensor into a $\chi d^2\times\chi d$ matrix
$M$. (iii) Perform a singular value decomposition (SVD) to the
matrix $M$. After truncating and reshaping, we get the tensors
$M_1$, $V_1$ and the updated diagonal matrix ${\tilde\lambda}_B$.
(iv) Insert the identity resolution
$\textbf{\textit{I}}=\lambda_C^{-1}\lambda_C$ on the right hand side
and contract the tensors $V_1$ and $\lambda_C^{-1}$to get the
updated ${\tilde\Gamma}_C$, as done in step (v). (vi) Contract $M_1$
and ${\tilde\lambda}_B$ into a matrix and perform a SVD again, one
gets the tensors $V_2$, $V_3$ and the updated ${\tilde\lambda}_A$.
(vii) Insert the identity resolution; (viii) All the tensors
${\tilde\Gamma}_A$, ${\tilde\Gamma}_B$, ${\tilde\Gamma}_C$, and the
diagonal matrices ${\tilde\lambda}_A$ and ${\tilde\lambda}_B$ are
updated. Shifting the action of the three-site gate $U_3$ one site
each time and repeating four times, we are able to update the
tensors under the imaginary time evolution of an slice $\delta\tau$.
Repeat the procedure until the ground-state energy per lattice
converges, an approximate ground-state wave function is generated in
the iMPS representation.

\section{Model}
Consider a spin 1/2 Heisenberg chain with the nearest-neighbor
coupling $J$ and next-to-the-nearest-neighbor coupling $J\delta$. It
is described by the Hamiltonian
\begin{equation}
  H=J \sum_{i=-\infty}^\infty \left(\textbf{S}^{[i]}\cdot \textbf{S}^{[i+1]}+
   \delta  ~\textbf{S}^{[i]}\cdot \textbf{S}^{[i+2]}\right), \label{ham2}
\end{equation}
where $\textbf{S}^{[i]}$ are the spin-$1/2$ Pauli operators at the
$i$-th site. The system is equivalent to a zigzag chain as shown in
Fig.~\ref{Zigzag}. We set the nearest-neighbor antiferromagnetic
coupling $J=1$ as the energy scale and consider the ratio interval
$0\leq \delta \leq 0.5$.  The system undergoes the KT transition at
$\delta_c \sim 0.2411$~\cite{value}: the Luttinger liquid state for
$\delta < \delta_c$ and the dimerized state for $\delta
> \delta_c$~\cite{dimer}, respectively. Actually, the KT transition accompanies
 a discrete $Z_2$ SSB in the dimerized phase.

\section {Ground-state fidelity per lattice site}
\begin{figure}
  \begin{center}
\includegraphics[width=2.9in]{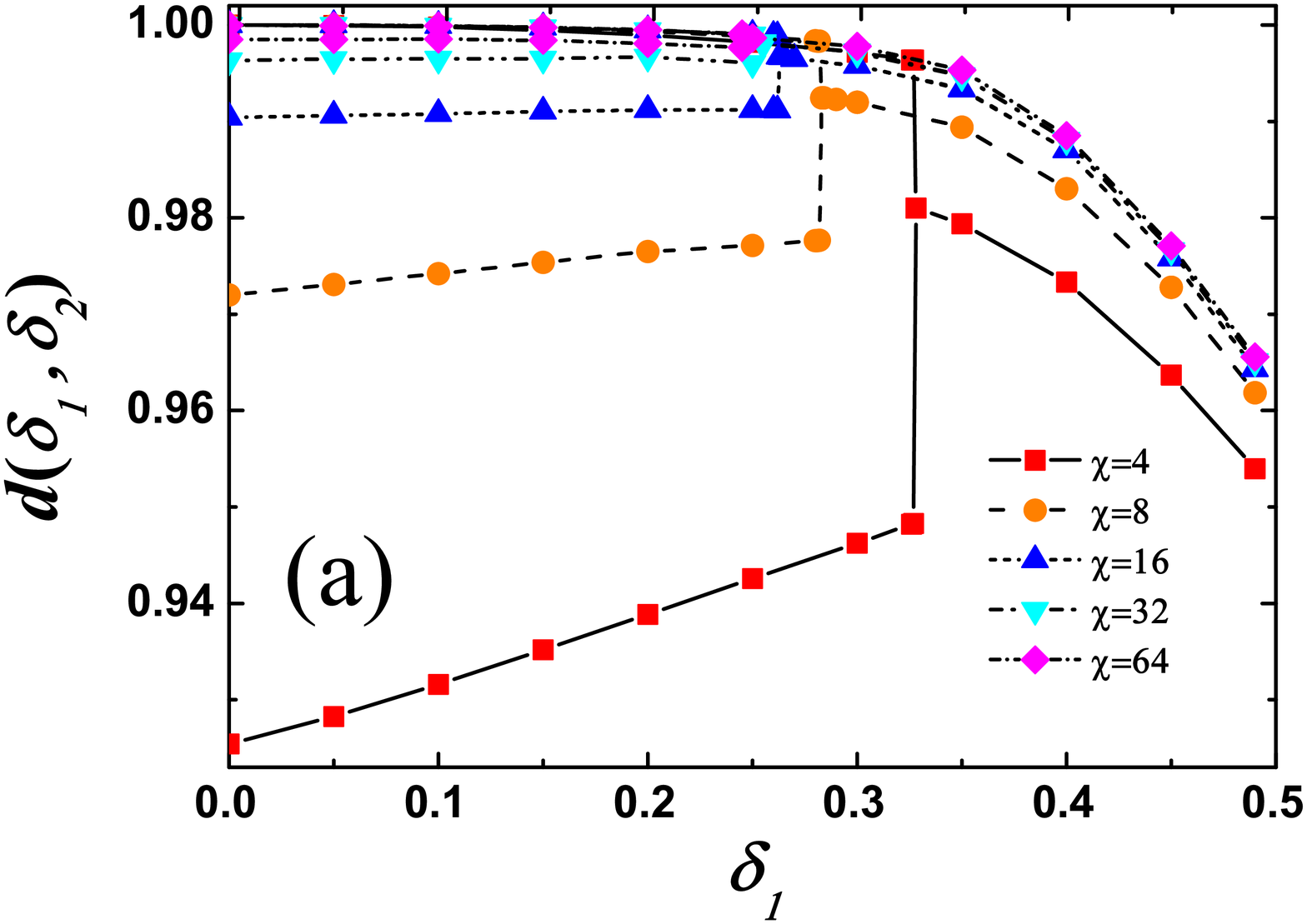}
\includegraphics[width=2.9in]{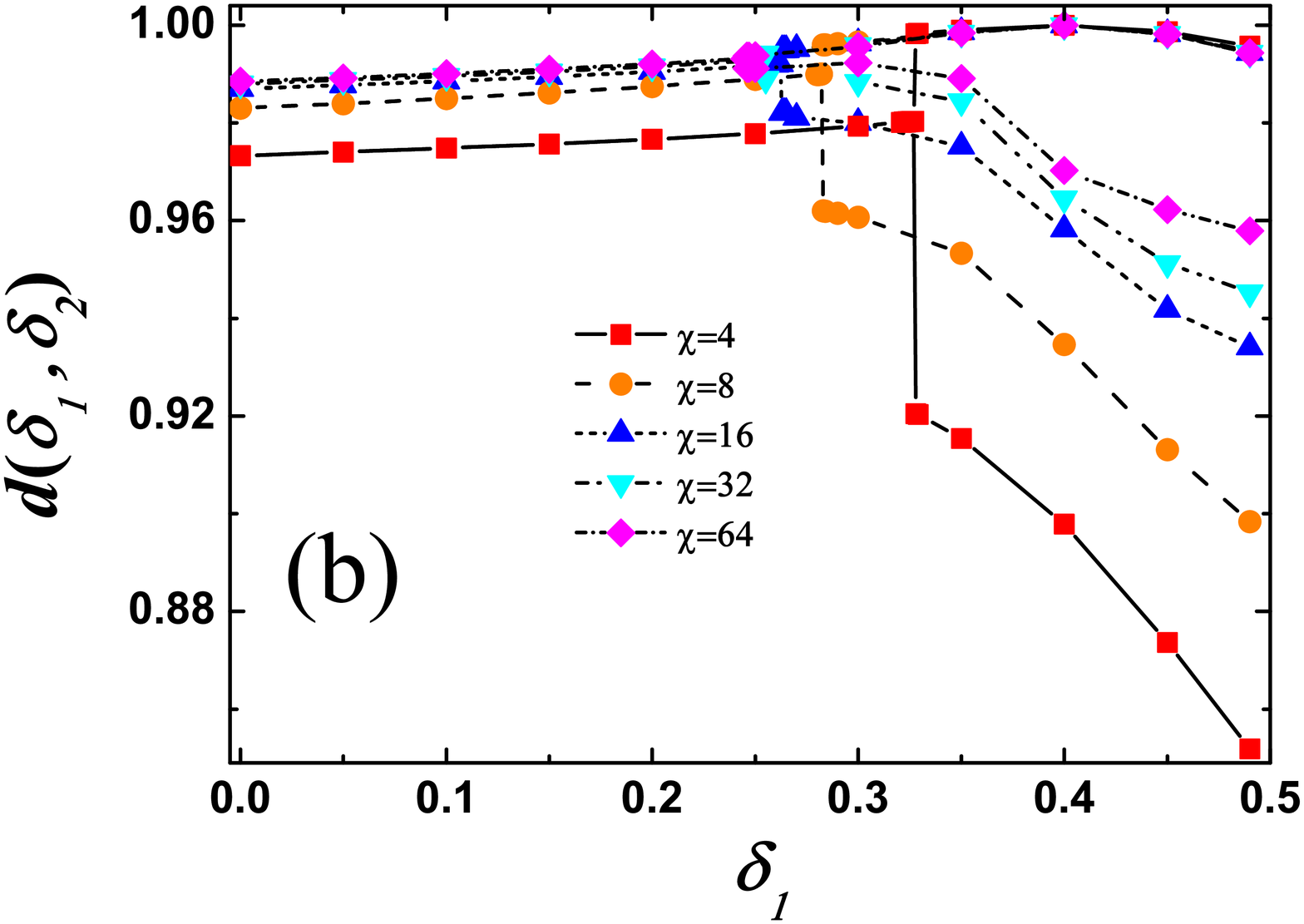}
  \end{center}
\caption{(Color online) The ground-state fidelity per lattice site,
$d(\delta_{1},\delta_{2})$, for the
 spin $1/2$ Heisenberg chain with the NNN interaction.
 We have chosen $\Psi(\delta_{2})$ as a reference
 state, with $\delta_{2}$ in different phases:
(a) $\delta_2=0$ in the $SU(2)$ symmetry-broken phase. There is a
bifurcation point in $d(\delta_{1},\delta_{2})$, which tends to
disappear, when $\chi$ approaches $\infty$.  (b) $\delta_2=0.4$ in
the $Z_{2}$ symmetric-broken phase. A bifurcation point always
exists whatever the truncation dimension $\chi$ we choose, as it
should be for a discrete symmetry SSB.  Therefore, the ground-state
fidelity per lattice site, $d(\delta_{1},\delta_{2})$, is able to
distinguish degenerate ground states, with a (pseudo) critical
 point as a bifurcation point.}\label{Fidelity}
\end{figure}
The ground-state fidelity per lattice site, $d(\delta_1,\delta_2)$,
with $\delta$ being the control parameter, is defined as the scaling
parameter: $F(\delta_1,\delta_2)\sim d(\delta_1,\delta_2)^N$, with
$N$ the total number of the lattice sites, and $F(\delta_1,
\delta_2) \equiv |\langle \Psi (\delta_2) |\Psi (\delta_1) \rangle |
$ between two ground-state wave functions $|\Psi (\delta_1) \rangle$
and $|\Psi (\delta_2) \rangle$. It characterizes how fast the
fidelity $F(\delta_1, \delta_2)$ varies when the thermodynamic limit
is approached~\cite{zhou,zov}. In fact, the ground-state fidelity
per lattice site $d(\delta_1, \delta_2)$ may be regarded as a
partition function per site of a classical statistical vertex
lattice model defined on the same lattice. That explains why $
d(\delta_1, \delta_2)$ is able to detect QPTs~\cite{zhou3}. That is,
the ground-state fidelity per lattice site exhibits a singular
behavior when the control parameter $\delta$ crosses a transition
point $\delta_c$.

Fig.~\ref{Fidelity} shows the ground-state fidelity per lattice
site, $d(\delta_{1},\delta_{2})$, as a function of $\delta_1$, when
$\delta_2$ is fixed, for the Heisenberg model with the NNN
interaction, for different values of the truncation dimension
$\chi$. The iMPS simulation is performed for a randomly chosen
initial state. It automatically induces degenerate ground-state wave
functions, which break the $SU(2)$ symmetry in the Luttinger liquid
phase and the $Z_2$ symmetry in the dimer phase.

In Fig.~\ref{Fidelity}(a), the reference state $\Psi(\delta_{2}=0)$
is chosen in the $SU(2)$ symmetry-broken phase. The ground-state
fidelity per lattice site, $d(\delta_{1},\delta_{2})$, takes
infinitely many values lying between two extreme values, as a
consequence of degenerate ground states, whereas in the $Z_2$
symmetry-broken phase, $d(\delta_{1},\delta_{2})$ yields just one
value. A bifurcation point occurs, which coincides with the
pseudo-transition point $\delta_c^\chi$. With increasing $\chi$, the
difference between two extreme values of $d(\delta_{1},\delta_{2})$
decreases. Note that, the bifurcation point tends to disappear when
$\chi$ goes to infinity,  implying that degenerate ground states
arise from the finiteness of the truncation dimension, an artifact
of the iMPS algorithm. In fact, all the degenerate ground states
should collapse into the genuine ground state as $\chi$ approaches
$\infty$, as required to keep consistent with the Mermin-Wagner
theorem: no continuous symmetry is spontaneously broken in
one-dimensional quantum systems. However, the critical point
$\delta_c$ may be determined by performing an extrapolation with
respect to a few reasonably small $\chi$'s.

In  Fig.~\ref{Fidelity}(b), the reference state
$\Psi(\delta_{2}=0.4)$ is chosen in the $Z_2$ symmetry-broken phase.
Bifurcation points occur again in the ground-state fidelity per
lattice site. However, they never vanish, when $\chi$ approaches
$\infty$. Instead, they tend to saturate when $\chi$ increases. This
is expected, since bifurcation points arise from the discrete group
$Z_2$ SSB.

Therefore, we conclude that, for a finite truncation dimension
$\chi$, the ground-state fidelity per lattice site,
$d(\delta_{1},\delta_{2})$, is able to distinguish degenerate
ground-state wave functions with a reference state in the
symmetry-broken phase. In addition, the iMPS algorithm enables us to
locate the KT transition point accurately by computing the
ground-state fidelity per lattice site, $d(\delta_{1},\delta_{2})$,
with a moderate computational cost.

\section{ Pseudo order parameter}
\begin{figure}
  \begin{center}
\includegraphics[width=2.8in]{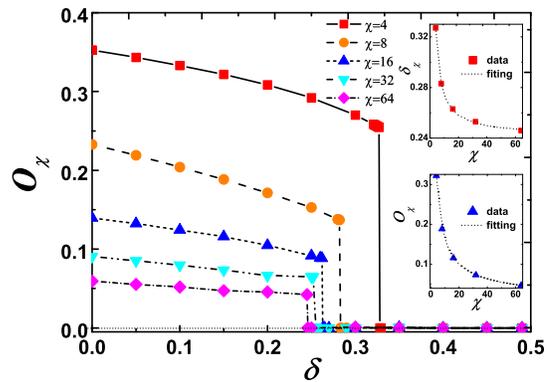}
  \end{center}
\caption{(Color online) The pseudo-local-order parameter $ O_{\chi}$
as a function of  $\delta$. The iMPS simulation is performed for a
randomly chosen initial state. An extrapolation of the pseudo
critical point $\delta_c^\chi$ is performed for pseudo-transition
points, at which the pseudo-order parameter become zero, yielding
the KT transition point $\delta_c=0.2418$ in the right-top inset.
The right-down inset shows that, in the critical phase, the
pseudo-order parameter is scaled down to zero according to
$O_{\chi}=a\chi^{-b}(1+c\chi^{-1})$, with a=0.5993, b=0.6150,
c=1.0379, to keep consistent with the Mermin-Wagner
theorem.}\label{order}
\end{figure}
As is well known, the iMPS algorithm yields the best approximation
to a ground-state wave function for a gapful lattice system.
However, for a gapless system with continuous symmetry, if the
truncation dimension $\chi$ is finite, the iMPS algorithm
automatically produces infinitely many degenerate ground states from
a randomly chosen initial state~\cite{zh}, each of which breaks the
continuous symmetry. That is, a numerical phenomenon occurs, which
shares all the features of an SSB resulting from random
perturbations. Such a pseudo-symmetry-broken order may be quantified
by introducing a pseudo-local-order parameter, which may be read off
from a reduced density matrix on a local area~\cite{zhou3}. However,
this phenomenon is in contradiction with the Mermin-Wagner theorem,
which states that no continuous symmetry is spontaneously broken for
quantum systems in one spatial dimension~\cite{mw}. To resolve this
apparent contradiction, one has to require that the
pseudo-local-order parameter must be scaled down to zero, when the
truncation dimension $\chi$ goes to $\infty$.  We emphasize that
both \textit{pseudo} SSB and  \textit{pseudo}-local-order parameter
arise from the artifact of the iMPS algorithm, in sharp contrast
with a genuine SSB and a local order parameter.

In the Heisenberg model with the NNN interaction, the
pseudo-local-order parameter may be chosen as $ O_\chi$=
$\sqrt{\langle S_{x}^{i}\rangle^2+\langle S_{y}^{i}\rangle^2+\langle
S_{z}^{i}\rangle^2}$ for the truncation dimension $\chi$.  The
pseudo-local-order parameter $ O_\chi$ is plotted as a function of
$\delta$ in Fig.~\ref{order}. The iMPS simulation is performed for a
randomly chosen initial state, with the truncation dimension $\chi$
to be 4, 8, 16, 32, and 64, respectively. Notice that $\hat{O}_\chi$
is zero in the $Z_2$ symmetry-broken phase, but nonzero in the
pseudo $SU(2)$ symmetry-broken phase, due to the finiteness of the
truncation dimension $\chi$.  However, this is nothing but an
artifact of the iMPS algorithm. Remarkably, one may take advantage
of this artifact to locate a critical point $\delta_c^\chi$. Note
that an extrapolation  with respect to $\chi$ may be performed for
the pseudo-phase-transition points $\delta_c^\chi$, at which the
pseudo-local-order parameter becomes zero. In the right-top inset of
Fig.~\ref{order}, the KT transition point $\delta_c=0.2418$ is
determined from such an extrapolation, which is comparable with
$\delta_c=0.2411$ from the level spectroscopy analysis~\cite{value}.
Here, we require that the pseudo-local-order parameter must vanish,
when $\chi$ goes to $\infty$, to keep consistent with the
Mermin-Wagner theorem. Therefore, a fitting function $
O_{\chi}=a\chi^{-b}(1+c\chi^{-1})$ is chosen, with a=0.5993,
b=0.6150, c=1.0379, as shown in the right-down inset in
Fig.~\ref{order}. This ensures that, when $\chi\rightarrow\infty$,
all the degenerate ground states collapse into the genuine ground
state.

\section{Conclusion}
We have investigated the KT transition for the spin $1/2$ Heisenberg
chain with the NNN interaction in the context of an iMPS algorithm,
a generalization of the infinite time-evolving block decimation
algorithm  to accommodate both the next-nearest-neighbor interaction
and spontaneous dimerization. It is demonstrated that, in the
critical regime, the algorithm automatically leads to infinitely
many degenerate ground-state wave functions, due to the finiteness
of the truncation dimension. This results in \textit{pseudo}
symmetry spontaneous breakdown, as reflected in a bifurcation in the
ground-state fidelity per lattice site. In addition, this allows to
introduce a pseudo-local-order parameter to characterize the KT
transition, which scales down to zero when $\chi$ approaches
infinity.

\section*{ Acknowledgements} The work is supported by the National
Natural Science Foundation of China (Grant No: 10874252) and the
Fundamental Research Funds for Central Universities (Project No.
CDJXS11102213).



\begin{thebibliography}{10}

\bibitem{qpt} S. Sachdev, {\it Quantum Phase Transition} (Cambridge University Press, Cambridge, 1999).

\bibitem{anderson} P. W. Anderson, {\it Basic Notions of Condensed Matter
Physics, Addison-Wesley: The Advanced Book Program} (Addison-Wesley,
Reading, MA, 1997).


\bibitem{coleman} S. Coleman, {\it An Introduction to Spontaneous Symmetry
Breakdown and Gauge Fields: Laws of Hadronic Matter},  ed. A.
Zichichi (Academic, New York, 1975).


\bibitem{wen} X.-G. Wen, {\it Quantum Field Theory of Many-Body Systems} (Oxford
University Press, Oxford, 2004).


\bibitem{kt} J. M. Kosterlitz and D. J. Thouless, J. Phys. C \textbf{ 6}, 1181
(1973); J. M. Kosterlitz, J. Phys. C \textbf{ 7}, 1046 (1974).


\bibitem{fss}M. N. Barber, {\it Phase Transitions and Critical Phenomena}, ed.
C. Domb and J. L. Lebowitz (Academic, London, 1983), Vol. 8.


\bibitem{np} J. S\'{o}lyom and T. A. L. Ziman, Phys. Rev. B \textbf{30}, 3980
(1984); R. G. Edwards, J. Goodman, and A. D. Sokal, Nucl. Phys. B
\textbf{354}, 289 (1991).

 \bibitem{zp} P. Zanardi and N.
Paunkovi\'{c}, Phys. Rev. E  \textbf{74}, 031123 (2006).

 \bibitem{zhou} H.-Q. Zhou and J. P. Barjaktarevi$\check{\rm c}$,
J. Phys. A: Math. Theor. \textbf{41}, 412001 (2008); H.-Q. Zhou,
J.-H. Zhao, and B. Li, J. Phys. A: Math. Theor. \textbf{41}, 492002
(2008); H.-Q. Zhou, arXiv: 0704.2945.

\bibitem{zov} H.-Q. Zhou, R. Or\'{u}s, and G. Vidal,
Phys. Rev. Lett. \textbf{100}, 080601 (2008).

\bibitem{fidelity1} P. Zanardi, M. Cozzini, and P. Giorda, J. Stat. Mech. L02002, (2007);
N. Oelkers and J. Links, Phys. Rev. B \textbf{75}, 115119 (2007); M.
Cozzini, R. Ionicioiu, and P. Zanardi, Phys. Rev. B \textbf{76},
104420 (2007); L. Campos Venuti and P. Zanardi, Phys. Rev. Lett.
\textbf{99}, 095701 (2007).

\bibitem{fidelity2}
 W.-L. You, Y.-W. Li, and S.-J. Gu, Phys. Rev. E \textbf{76},
022101 (2007); S. J. Gu, H. M. Kwok, W. Q. Ning, and H. Q. Lin,
Phys. Rev. B \textbf{77}, 245109 (2008); M. F. Yang, Phys. Rev. B
\textbf{76}, 180403(R) (2007);  Y. C. Tzeng and M. F. Yang, Phys.
Rev. A \textbf{77}, 012311 (2008); J. O. Fj{\ae}restad, J. Stat.
Mech.: Theory Exp. (2008) P07011; T. Liu, Y.-Y. Zhang, Q.-H. Chen,
and K.-L. Wang, Phys. Rev. A \textbf{80}, 023810 (2009); J. Sirker,
Phys. Rev. Lett. \textbf{105}, 117203 (2010).

\bibitem{fidelity3} S. Chen, L. Wang, Y. Hao, and Y. Wang, Phys. Rev. A \textbf{77}, 032111 (2008);
S. Chen, L. Wang, S.-J. Gu, and Y. Wang, Phys. Rev. E \textbf{76},
061108 (2007).

\bibitem{rams} M. M. Rams and B. Damski, Phys. Rev. Lett. \textbf{106}, 055701 (2011).

\bibitem{vidal1} G. Vidal, Phys. Rev. Lett. \textbf{98}, 070201 (2007).

\bibitem{PEPS} F. Verstraete, and J. I. Cirac, e-print arXiv:cond-mat/0407066.

\bibitem{zh} H.-L. Wang, J.-H. Zhao, B. Li, and H.-Q. Zhou, arXiv: 0902.1670.


\bibitem{mw} N. D. Mermin and H. Wagner, Phys. Rev. Lett.  \textbf{17}, 1133 (1966).


\bibitem{suzuki} M. Suzuki, Phys. Lett. A \textbf{146}, 319 (1990).

\bibitem{value} K. Okamoto and K. Nomura, Phys. Lett. A \textbf{169}, 433 (1992).
\bibitem{dimer} F. D. M. Haldane, Phys. Rev. B \textbf{25}, 4925 (1982).

\bibitem{zhou3} H.-Q. Zhou, arXiv: 0803.0585.




 \end{thebibliography}
\end{document}